\documentclass{article}
\usepackage{epsfig}

\newcommand{\be}{\begin{equation}}
\newcommand{\ee}{\end{equation}}
\newcommand{\ba}{\begin{eqnarray}}
\newcommand{\ea}{\end{eqnarray}}

\input{tcilatex}
\begin{document}

\title{Improved Quantum Hard-Sphere Ground-State Equations of State}
\author{M.A. Sol\'{\i}s$^{1,2}$, M. de Llano$^{3}$, J.W. Clark$^{1}$, and
George A. Baker, Jr.$^{4}$ \\
\vspace{-0.2cm} $^{1}$ Department of Physics, Washington University \\
St. Louis, Missouri 63130, USA \\
\vspace{-0.2cm} $^{2}$ Instituto de F\'{\i}sica, Universidad Nacional Aut%
\'{o}noma de M\'{e}xico \\
Apdo. Postal 20-364, 01000 M\'{e}xico, D.F., Mexico \\
\vspace{-0.2cm} $^{3}$ Instituto de Investigaciones en Materiales,
Universidad Nacional Aut\'{o}noma de M\'{e}xico, \\
\vspace{-0.2cm}Apdo. Postal 70-360, 04510 M\'{e}xico, DF, Mexico\\
\vspace{-0.2cm} \& \\
\vspace{-0.2cm} Consortium of the Americas for Interdisciplinary Science,\\
University of New Mexico, Albuquerque, NM 87131, USA\\
\vspace{-0.2cm} $^{4}$Theoretical Division, University of California, \\
Los Alamos National Laboratory, Los Alamos, NM 87545, USA \\
}
\maketitle

\begin{abstract}
The London ground-state energy formula as a function of number density for a
system of identical boson hard spheres, corrected for the reduced mass of a
pair of particles in a \textquotedblleft
sphere-of-influence\textquotedblright\ picture, and generalized to fermion
hard-sphere systems with two and four intrinsic degrees of freedom, has a
double-pole at the ultimate \textit{regular} (or periodic, e.g.,
face-centered-cubic) close-packing density usually associated with a
crystalline branch. Improved fluid branches are contructed based upon exact,
field-theoretic perturbation-theory low-density expansions for many-boson
and many-fermion systems, extrapolated to intermediate densities via Pad\'{e}
and other approximants, but whose ultimate density is irregular or \textit{%
random} closest close-packing as suggested in studies of a classical system
of hard spheres. Results show substantially improved agreement with the best
available Green-function Monte Carlo and diffusion Monte Carlo simulations
for bosons, as well as with ladder, variational Fermi hypernetted chain, and
so-called L-expansion data for two-component fermions.

\newline
\end{abstract}


\section{Introduction}

An analytical formula for the ground-state energy $E$ of an $N$%
-hard-sphere-boson system of volume $\Omega $ for all particle-number
densities $\rho \equiv N/\Omega $ was proposed by London \cite{London} as 
\begin{equation}
E/N=\frac{2\pi \hbar ^{2}c}{m}\frac{1}{(\rho ^{-1/3}-\rho _{0}^{-1/3})^{2}}%
\frac{1}{(\rho ^{-1/3}+b\,\rho _{0}^{-1/3})}  \label{london}
\end{equation}%
where $m$ is the particle mass, $c$ is the hard-sphere diameter and the
constant $b$ equals $2^{5/2}/\pi -1$.\ Here, $\rho _{0}\equiv \sqrt{2}/c^{3}$
is the assumed ultimate regular (or periodic) close-packing density at which
a system of identical classical hard spheres close-pack in a
primitive-hexagonal arrangement, e.g., face-centered-cubic or hexagonal. As
remarked by Rogers \cite{Rogers} this is what \textquotedblleft many
mathematicians believe and all physicists know\textquotedblright\ to be the
case. However, the Kepler 1611 conjecture \cite{Rogers1}\ that $\rho
_{0}\equiv \sqrt{2}/c^{3}$ is the ultimate packing density for identical
hard spheres seems to be approaching theorem status \cite{Hales} after many
attempts of proof.

The justification given for (\ref{london}) is that it reduces smoothly to
limiting expressions at both low and high densities, namely 
\begin{equation}
E/N\;\smash {\ \mathop{\relbar\joinrel\longrightarrow}\limits_{\rho \to 0}\
\ }\;(2\pi \hbar ^{2}/m)\rho c  \label{tocero}
\end{equation}%
\begin{equation}
E/N\;\smash {\ \mathop{\relbar\joinrel\longrightarrow}\limits_{\rho \to
\rho_0}\ \ }\;A(\hbar ^{2}/2m)(\rho ^{-1/3}-\rho _{0}^{-1/3})^{-2}
\label{tocp}
\end{equation}%
but gives no indication of a \textquotedblleft freezing\textquotedblright\
or Kirkwood \cite{Kirw}\ phase transition at some number density $\rho $
between $0$ and $\rho _{0}$. Here $A=\pi ^{2}/2^{1/3}\simeq 7.8335$ is a
constant called the \textit{residue }of the second-order (or double) pole at
close packing. Using the polyhedron cell method suggested in Ref. \cite{Yang}%
, the value of $A$ has been predicted \cite{Hubbard}\ theoretically to lie
within the rigorous range 
\begin{equation}
1.63\leq A\leq 27.0\,.  \label{rigorousRange}
\end{equation}%
The low-density leading term (\ref{tocero}) is the celebrated Lenz \cite%
{Lenz} term, calculated by him as the leading correction to the energy
arising from an \textquotedblleft excluded volume\textquotedblright\ effect.
The Lenz term has finally been rigorously established \cite{Lieb98}. The
limit (\ref{tocp}) comes from the lowest Schr\"{o}dinger equation eigenvalue
of a particle in a spherical cavity, and is just the kinetic energy of a
point particle of mass $m$ inside the cavity of radius $r-c$, where $r$ is
the average separation between two neighboring hard spheres and $r=(\sqrt{2}%
/\rho )^{1/3}$ by assuming a primitive-hexagonal packing arrangement for the
cavities.

More recently it was found \cite{solis1}, however, that the arguments
leading to the high-density limit of the original \cite{London} (boson)
London equation (\ref{london}) are flawed by a\textit{\ }fundamental error:
the spherical cavity of radius $r-c$ alluded to above in reality refers to
the \textquotedblleft sphere of influence\textquotedblright\ of \textit{two}
particles. Thus, the particle mass used in obtaining (\ref{tocp}) should
refer to the\textit{\ reduced mass} $m/2$ of the pair. This yields the
constant 
\begin{equation}
b\equiv 2^{3/2}/\pi -1  \label{bnew}
\end{equation}
instead of the constant\thinspace\ $2^{5/2}/\pi -1$ \thinspace given by
London for (\ref{london}). The result (\ref{london}) with (\ref{bnew}) is
designated the \textit{modified London (ML) equation}. It continues to
satisfy (\ref{tocero}) as this is independent of the constant $b$ but the
residue {$A$ in (\ref{tocp}) now becomes $2^{2/3}\pi ^{2}\simeq 15.667$ }%
instead of the previous $\pi ^{2}/2^{1/3}\simeq 7.8335$\ associated with the
original London equation, and fully agrees with the empirical residue of $%
15.7$ $\pm $ $0.6$, extracted by Cole \cite{Cole} from high-pressure
crystalline-branch data in $^{3}$He, $^{4}$He, H$_{2}$ and D$_{2}$ systems.
Moreover, this ML equation exhibits dramatically better agreement than the
original London (L) equation with Green-function Monte Carlo (GFMC) \cite%
{GFMC} computer-simulation data points for both fluid and crystalline
branches of the boson hard-sphere system.

A generalized London equation has also been proposed \cite{Ren} for $N$-%
\textit{fermion} hard-sphere systems with $\nu $ intrinsic degrees of
freedom for each fermion. Here $\nu =2$ for, say, liquid $^{3}$He or neutron
matter, both constituent fermions of which have spin $1/2$, and $\nu =4$ for
nuclear matter consisting of both neutrons \textit{and} protons of spin $1/2$%
. As $\nu $ is essentially the maximum occupation in a given single-particle
quantum state, it can be taken as infinite in the case of bosons. For
fermions, two differences appear with respect to the boson London Formula:
(a) unlike the boson case, the ground-state kinetic energy for fermions is
nonzero and is added as a well-known \cite{FW} $\nu $-dependent leading
term; and (b) the constant $b$ is allowed to be $\nu $-dependent, being
replaced by 
\begin{equation}
b_{\nu }(\nu )=[(\nu -1)/\nu ](b+1)-1  \label{bnu}
\end{equation}

which clearly approaches $b$\ as $\nu \rightarrow \infty $. The latter form
also ensures a $\nu $-\textit{independent} energy at close-packing where,
since the spheres can be labeled so that indistinguishability as well as
particle statistics disappears, as expected in this classical limit.
Substitution of $b_{\nu }$ for the constant $b$ in (\ref{london}) gives a
generalized form of the modified London equation (ML$_{\nu }$) 
\begin{equation}
E/N=C_{\nu }\,\rho ^{2/3}+\bigg(\frac{\nu -1}{\nu }\bigg)\frac{2\pi \hbar
^{2}c}{m}\frac{1}{(\rho ^{-1/3}-\rho _{0}^{-1/3})^{2}}\frac{1}{[\rho
^{-1/3}+b_{\nu }(\nu )\,\rho _{0}^{-1/3}]}  \label{gen}
\end{equation}%
with 
\begin{equation}
C_{\nu }\equiv \frac{3\hbar ^{2}}{10m}\bigg(\frac{6\pi ^{2}}{\nu }\bigg)%
^{2/3}\smash {\ \mathop{\relbar\joinrel\longrightarrow}\limits_{\nu \to
\infty}\ \ }0.
\end{equation}%
For $\nu \rightarrow \infty $, $b(\nu )\rightarrow b$ according to (6), and (%
\ref{gen}) goes over into the boson case (\ref{london}) 
because $C_{\nu }$ vanishes in this limit. The low-density limit of (\ref%
{gen}) is 
\begin{equation}
E/N\smash {\ \mathop{\relbar\joinrel\longrightarrow}\limits_{\rho \to 0}\ \ }%
C_{\nu }\,\rho ^{2/3}+\bigg(\frac{\nu -1}{\nu }\bigg)\frac{2\pi \hbar ^{2}}{m%
}\rho c
\end{equation}%
where the second term on the rhs is the Lenz term for $\nu $-component
fermions in 3D. On the other hand, for $\rho \rightarrow \rho _{0}\equiv 
\sqrt{2}/c^{3}$ one sees that (\ref{gen}) reduces to (\ref{tocp}) as it
should. In other words, hard-sphere fermions, bosons or \textquotedblleft
boltzons\textquotedblright\ must all close-pack regularly\textit{\ at the
same density}. From this it follows that the residue for bosons or fermions
is the same and equal to $2^{2/3}\,\pi ^{2}\simeq 15.667$, in excellent
agreement with the empirical Ref.~\cite{Cole} value of $15.7$ $\pm $ $0.6$.

For bosons, in addition to the Lenz term (\ref{tocero}) for the low-density
fluid branch, several higher-order corrections to the ground-state energy
per particle have been derived using quantum field-theoretic many-boson
perturbation theory \cite{boson}\cite{boson1}. They give 
\begin{equation}
E/N=\frac{2\pi \hbar ^{2}\rho c}{m}\bigg\{1+C_{1}(\rho
c^{3})^{1/2}+C_{2}\rho c^{3}\,\ln (\rho c^{3})+C_{3}\rho c^{3}+o(\rho c^{3})%
\bigg\}\,  \label{bos}
\end{equation}
for $\rho c^{3}\ll 1$, where $C_{1}=128/15\sqrt{\pi }$ and $C_{2}=8(4\pi /{3}%
-\sqrt{3})$, but $C_{3}$ is an as yet unknown constant. Here, $c$ denotes
the S-wave scattering length for a general potential; for a hard-core
potential it is just the hard-sphere diameter. The series is clearly not a 
\textit{pure} power series expansion, and is at best an asymptotic series.

Similarly, for an $N$-fermion hard-sphere system the corresponding series is 
\cite{fermion} 
\begin{eqnarray}
E/N &=&\frac{3}{5}\frac{\hbar ^{2}k_{F}^{2}}{2m}\bigg\{%
1+C_{1}(k_{F}c)+C_{2}(k_{F}c)^{2}  \nonumber \\
&+&[C_{3}r_{0}/2c+C_{4}A_{1}(0)/c^{3}+C_{5}](k_{F}c)^{3}+C_{6}(k_{F}c)^{4}%
\ln (k_{F}c)  \nonumber \\
&+&[C_{7}r_{0}/2c+C_{8}A_{0}^{\prime \prime
}(0)/c^{3}+C_{9}](k_{F}c)^{4}+o(k_{F}c)^{4}\bigg\}  \label{fer}
\end{eqnarray}
for $k_{F}c\ll 1$ and\ where the $C_{j}$ ($j=1,2,...,9$) are dimensionless
coefficients depending on $\nu $; they are given in Ref.~\cite{coeff} for $%
\nu =2$ and $\nu =4$. The Fermi momentum $\hbar k_{F}$ is defined through
the fermion-number density 
\begin{equation}
\rho \equiv N/\Omega =\nu k_{F}^{3}/6\pi ^{2}  \label{density}
\end{equation}
with $\Omega $ the system volume, so that the Lenz term expressed in terms
of $\rho $ is identical to the boson Lenz term apart from a factor of $(\nu
-1)/\nu $ which is the average number of fermions the Pauli Principle allows
a given fermion to interact with at the shortest possible range.

Unfortunately, both low-density expansions (\ref{bos}) and (\ref{fer}) lack
accuracy at moderate to high densities, including the saturation (or
equilibrium, zero-pressure) densities of liquid $^{4}$He ($\nu =\infty $) 
\cite{Pineda}\ and liquid $^{3}$He ($\nu =2$) or nuclear matter ($\nu =4$).
However, one can extrapolate the series for hard-sphere systems to physical
and even to close-packing densities through the use of Pad\'{e} \cite%
{BakerGraves-Morris}\ and/or a modest extension of these called the
\textquotedblleft tailing\textquotedblright\ \cite{tailing} approximants.
The so-called quantum thermodynamic (or van der Waals) perturbation theory
(QTPT) \cite{QTPT,QTPT2} has provided fairly accurate representations of the 
\textit{fluid} branch of the equation of state of quantum hard-sphere
systems \cite{Annals}, even beyond freezing (or, Kirkwood) phase transition
densities, but without sufficient credibility as one approaches close
packing. This is clear since one does not possess a \textit{single }%
ground-state energy function with implicit information of \textit{both}
fluid and crystalline branches, with presumably different close-packing
ultimate densities.

In Section 2 we discuss the double- (or second-order-) pole behavior for the
equation-of-state fluid branch conceivably ending at \textit{random} closest
close-packing, instead of the \textit{regular} close-packing at which the
crystalline branch terminates; in Secs. 3 and 4 we construct analytical
expressions for the fluid branches for hard-sphere bosons and fermions,
respectively. Sec. 5 gives our conclusions.

\section{Double-pole conditions at close-packing}

We shall assume that the fluid branch of the hard-sphere equation of state
will terminate not at the regular close-packing density $\rho _{0}$ but
rather\ at the random closest close packing (rccp), sometimes called the
Bernal, density $\rho _{rccp}$ (or \textquotedblleft maximally random
jammed\textquotedblright\ packing \cite{Torquato}). Its value was originally
determined empirically \cite{ScottKilgour}\ with actual ball-bearing
packings. Near the density $\rho _{rccp}$ we expect, based on (\ref{tocp}),
that the energy for a hard sphere boson or fermion gas has the following
behavior 
\begin{equation}
E/N\;\smash {\ \mathop{\relbar\joinrel\longrightarrow}\limits_{\rho \to
\rho_{rccp}}\ \ }\;A(\hbar ^{2}/2m)(\rho ^{-1/3}-\rho _{rccp}^{-1/3})^{-2}
\label{C1}
\end{equation}%
with $A$ the \textit{residue} which could be different for each system.\
Random close-packing densities range \cite{Jaeger}\ from about $0.06\rho
_{0} $ to $0.86\rho _{0}\equiv \rho _{rccp}$.

The derivative of (\ref{C1})\ with respect to $\rho $ then tends
asymptotically to 
\begin{equation}
\frac{d(E/N)}{d\rho }\;\smash {\
\mathop{\relbar\joinrel\longrightarrow}\limits_{\rho \to \rho_{rccp}}\ \ }\;%
\frac{A(2/3)(\hbar ^{2}/2m)}{(\rho ^{-1/3}-\rho _{rccp}^{-1/3})^{3}\rho
^{4/3}}  \label{C2}
\end{equation}%
while 
\begin{equation}
\frac{d\ln (E/N)}{d\rho }\;\smash {\
\mathop{\relbar\joinrel\longrightarrow}\limits_{\rho \to \rho_{rccp}}\ \ }\;%
\frac{2/3}{(\rho ^{-1/3}-\rho _{rccp}^{-1/3})\rho ^{4/3}}  \label{C3}
\end{equation}%
is residue independent. We shall assume that $A$ is the same for boson as
for fermion hard spheres and that their rccp density is likewise identical
since at closest close-packing the particles become localized by definition,
enabling one to formally label each particle; this makes them
distinguishable thus rendering (quantum) statistics irrelevant. Note that
the pressure $P=\rho ^{2}[d(E/N)/d\rho ]$ from (\ref{C2}) also diverges as $%
\rho \longrightarrow \rho _{rccp}$, as expected.

\section{Boson hard-sphere fluid}

In order to extrapolate the low-density series (\ref{bos}) to higher
densities we start by writing it as 
\begin{equation}
E/N=\frac{2\pi \hbar ^{2}}{m}\rho ce_{0}(x)\quad  \label{bosons}
\end{equation}%
where $x\equiv (\rho c^{3})^{1/2}\,$and\ 
\begin{equation}
e_{0}(x)\equiv 1+C_{1}x+C_{2}x^{2}\,\ln x^{2}+C_{3}x^{2}+O(x^{3}\,\ln
x^{2})\,  \label{e0b.a}
\end{equation}%
for $x\ll 1$. Alternatively, one can rewrite this series as 
\begin{equation}
e_{0}^{-1/2}(x)=1+K_{1}\,x+K_{2}\,x^{2}\,\ln x^{2}+K_{3}\,x^{2}+O(x^{3}\,\ln
x^{2})  \label{e0b.b}
\end{equation}%
\begin{table}[b]
\begin{center}
\begin{tabular}{||c|c|c|c||}
\hline\hline
Bosons ($\nu = \infty$) & $i=$1 & 2 & 3 \\ \hline
$C_i$ & 4.81441778 & 19.65391518 & ``73.296" \\ \hline
$K_i$ & -2.40720889 & -9.826957589 & ``-27.956" \\ \hline\hline
\end{tabular}
\\[0pt]
\end{center}
\caption{Coefficients $C_{i}$ and $K_{i}$ for bosons appearing in (\protect
\ref{e0b.a}) and (\protect\ref{e0b.b}), respectively. Numbers in quotation
marks are determined as indicated in text.}
\label{tab:BosonCoef}
\end{table}
where the $K_{i}$'s are expressible in terms of the $C_{i}$'s. As $C_{3}$ is
to date unknown, consequently $K_{3}$ is also unknown. Values of the $C_{i}$%
's and $K_{i}$'s are given in Table~\ref{tab:BosonCoef}. We analyze the
series $e_{0}^{-1/2}(x)$ instead of the series $e_{0}(x)$ to ensure that any
zeros in its extrapolants, say $\epsilon _{0}^{-1/2}(x),$ are double (or
second-order) poles in the energy as one expects at any kind of close
packing. The extrapolants are generated as a quotient of two polynomials
such that on expansion one recovers the first terms of the original series.
Series (\ref{e0b.b}) with three terms beyond unity has twelve extrapolants
correctly generated in Ref. \cite{formas} but fitted there to erroneous
(i.e., to one-half the correct values) GFMC data points \cite{GFMC}.
Adjusting various extrapolants \cite{Annals} to best-fit the four known\
GFMC data points ensures a good value for the unknown coefficient $K_{3}$ in
(\ref{e0b.b}). The extrapolant labeled \textquotedblleft XI (bosons)" in
Fig. 2 of Ref. \cite{Annals}\ had the least mean-square deviation with
respect to the four GFMC fluid-branch data points. Therefore, we adopt it as
our best initial extrapolant. The ground-state energy per particle for boson
hard spheres was thus represented (symbol $\doteq $)\ by 
\begin{equation}
E/N\doteq \frac{2\pi \hbar ^{2}}{m}\rho c\epsilon _{0}(\rho )
\label{bosonEoverN}
\end{equation}%
with $K_{3}\simeq -27.956$. However, as diffusion Monte Carlo (DMC)
calculations became available \cite{Boronat}\ spanning a wider range of
densities in the fluid region than GFMC data, we realized that although our
expression XI$(x)$ in Eq. (17) of Ref. \cite{Annals}\ agrees well with DMC
and GFMC data around the freezing transition, its disagreement with the DMC
data at low to intermediate densities suggested the possibility of improving
the extrapolant. As will be seen, the new extrapolant $\epsilon _{0}^{-1/2}$%
\ predicts a random closet close-packing (rccp) density $\rho _{rccp}/\rho
_{0}\simeq 0.776$ which is only about $10\%$ below the classical
hard-spheres empirical \cite{ScottKilgour}\ rccp value $\simeq 0.86$
mentioned before and also assumed to be the ultimate rccp density for
quantum hard-sphere fluids.

In order to improve the fluid-branch expression of Ref. \cite{Annals}\ for 
\textit{low to} \textit{intermediate} densities we use the two double-pole
conditions (\ref{C1}) and (\ref{C2}) which lead to the following conditions
on the extrapolant $\epsilon _{0}(x)$ to be used in (\ref{bosonEoverN}),
namely 
\[
\epsilon _{0}=\frac{mE}{N2\pi \hbar ^{2}\rho c}\quad 
\smash {\
\mathop{\relbar\joinrel\longrightarrow}\limits_{\rho \to \rho_{rccp}}\ \ }%
\quad \frac{A}{4\pi \rho c}(\rho ^{-1/3}-\rho _{rccp}^{-1/3})^{-2}\,. 
\]%
This is equivalent to 
\begin{equation}
\epsilon _{0}^{-1/2}(\rho )\quad \smash {\
\mathop{\relbar\joinrel\longrightarrow}\limits_{\rho \to \rho_{rccp}}\ \ }%
\quad \left[ {A}/{4\pi \rho c}\right] ^{-1/2}(\rho ^{-1/3}-\rho
_{rccp}^{-1/3})\quad \smash {\
\mathop{\relbar\joinrel\longrightarrow}\limits_{\rho \to \rho_{rccp}}\ \ }%
\quad 0\,.  \label{C1a}
\end{equation}%
The condition (\ref{C2}) gives 
\begin{equation}
\frac{d(\epsilon _{0}^{-1/2})}{d\rho }\;\smash {\
\mathop{\relbar\joinrel\longrightarrow}\limits_{\rho \to \rho_{rccp}}\ \ }%
\;-(1/3)\left[ {A}/{4\pi c}\right] ^{-1/2}\rho _{rccp}^{-5/6}\,.  \label{C2a}
\end{equation}

Strictly, any log term should be accompanied by a constant, if known,
because the scaling of $\rho $ by $c^{3} $ is arbitrary. We thus propose the
representation of $e_{0}(x)$ in (\ref{e0b.a}) as given by 
\begin{equation}
e_{0}^{-1/2}(x)\doteq \frac{1+K_{1}\,x+\beta x^{2}+\gamma x^{3}}{%
1-K_{2}\,x^{2}\ln \,x^{2}+\alpha x^{2}}\equiv \epsilon _{0B}^{-1/2}(x)
\label{eps0B}
\end{equation}%
where $\alpha ,$ $\beta $ and $\gamma $ are to be determined from (\ref{C1a}%
) and (\ref{C2a})\ and by fitting both DMC \cite{Boronat} and GFMC (\cite%
{GFMC}, Table I) data. In this approximant the terms in $x^{2}\ln \,x$ and $%
x^{2}$ are kept together. Condition (\ref{C1a}) applied to (\ref{eps0B})
gives 
\begin{equation}
1+K_{1}\,x_{rccp}+\beta x_{rccp}^{2}+\gamma x_{rccp}^{3}=0\,.  \label{gamma}
\end{equation}%
The second condition (\ref{C2a}) can be rewritten as 
\begin{equation}
\left. \frac{d(\epsilon _{0B}^{-1/2})}{dx}\frac{dx}{d\rho }\right\vert
_{x=x_{rccp}}=\frac{K_{1}+2\beta x_{rccp}+3\gamma x_{rccp}^{2}}{%
1-K_{2}\,x_{rccp}^{2}\ln \,x_{rccp}^{2}+\alpha x_{rccp}^{2}}\frac{c^{3}}{%
2x_{rccp}}=-(1/3)\left[ {A}/{4\pi c}\right] ^{-1/2}\rho _{rccp}^{-5/6}\,.
\label{deps0-1/2}
\end{equation}%
Substituting (\ref{gamma}) in the last equation we obtain $\beta $ in terms
of $\alpha $, namely 
\begin{equation}
-\beta x_{rccp}^{2}=3+2K_{1}x_{rccp}-\left[ {A}/{4\pi }\right] ^{-1/2}\frac{2%
}{3}x_{rccp}^{1/3}(1-K_{2}\,x_{rccp}^{2}\ln \,x_{rccp}^{2}+\alpha
x_{rccp}^{2})\,.  \label{beta}
\end{equation}%
Now substituting (\ref{beta}) in (\ref{gamma}) we arrive at 
\begin{equation}
\gamma x_{rccp}^{3}=2+K_{1}x_{rccp}-\left[ {A}/{4\pi }\right] ^{-1/2}\frac{2%
}{3}x_{rccp}^{1/3}(1-K_{2}\,x_{rccp}^{2}\ln \,x_{rccp}^{2}+\alpha
x_{rccp}^{2})\,.  \label{gamma2}
\end{equation}%
Introducing (\ref{beta}) and (\ref{gamma2}) in (\ref{eps0B}), we get 
\begin{eqnarray}
\epsilon _{0B}^{-1/2}(x)=\left[ 1+K_{1}\,x+(x/x_{rccp})^{2}%
\{-3-2K_{1}x_{rccp}\}+(x/x_{rccp})^{3}\{2+K_{1}x_{rccp}\}+\left[ {A}/{4\pi }%
\right] ^{-1/2}\frac{2}{3}x_{rccp}^{1/3}\right. &&  \nonumber \\
\left. \times (1-K_{2}\,x_{rccp}^{2}\ln \,x_{rccp}^{2}+\alpha
x_{rccp}^{2})\{(x/x_{rccp})^{2}-(x/x_{rccp})^{3}\}\right] [1-K_{2}\,x^{2}\ln
\,x^{2}+\alpha x^{2}]^{-1} &&  \label{Y}
\end{eqnarray}%
from which after some algebra one obtains a single equation for $\alpha ,$
namely 
\begin{eqnarray}
&&\left[ \epsilon _{0B}^{-1/2}(\alpha ,A,x)\right] (1-K_{2}\,x^{2}\ln
\,x^{2})-1-K_{1}\,x+(x/x_{rccp})^{2}\{3+2K_{1}x_{rccp}\}  \nonumber \\
&&-\left[ {A}/{4\pi }\right] ^{-1/2}\frac{2}{3}x_{rccp}^{1/3}(1-K_{2}%
\,x_{rccp}^{2}\ln
\,x_{rccp}^{2})\{(x/x_{rccp})^{2}-(x/x_{rccp})^{3}\}-(x/x_{rccp})^{3}%
\{2+K_{1}x_{rccp}\}  \nonumber \\
&&\qquad \qquad \qquad \qquad \qquad =\alpha \{x_{rccp}^{2}\left[ {A}/{4\pi }%
\right] ^{-1/2}\frac{2}{3}%
x_{rccp}^{1/3}[(x/x_{rccp})^{2}-(x/x_{rccp})^{3}]-x^{2}[\epsilon
_{0B}^{-1/2}(\alpha ,A,x)]\}\,.  \label{alpha}
\end{eqnarray}%
where we have explicitly written the dependence of $\epsilon _{0B}^{-1/2}(x)$
on $\alpha $ and $A$. To determine $\alpha $ from the DMC \cite{Boronat}
and/or GFMC data we must calculate the values $\alpha _{i}^{DMC}$ [from (\ref%
{alpha}) after replacing $\epsilon _{0B}^{-1/2}(\alpha ,A,x)$ by the $%
\epsilon _{0-DMC}^{-1/2}(x_{i}^{DMC})$ obtained from (\ref{bosonEoverN}) as $%
(2\pi \hbar ^{2}\rho cN/Em)^{1/2}$ with $E/N$ the energy from DMC
calculations] for each $x_{i}^{DMC}$ for $i=1,2,\cdots N$ values, and then
minimizes $\sum_{i=1}^{N}(\alpha _{i}^{DMC}-\alpha )^{2}$ by imposing 
\[
\frac{d}{d\alpha }\sum_{i=1}^{N}(\alpha _{i}^{DMC}-\alpha )^{2}=0 
\]%
which gives 
\[
\alpha =\sum_{i=1}^{N}\alpha _{i}^{DMC}/N. 
\]%
Since the fluid branch GFMC data are a subset of DMC data, we have used
these to calculate $\alpha $ here, determining $A$ in the next step. For
residue $A$ 
fixed at $2^{2/3}\pi ^{2}\simeq 15.667$ as described below (\ref{bnew}), we
obtain an optimal $\alpha \simeq 114.282$ which from (\ref{beta}) and (\ref%
{gamma2})\ leads to $\beta \simeq 74.0891$ and $\gamma \simeq -65.9475$. The
curve then corresponding to (\ref{eps0B}) is labeled B1 in Fig. \ref%
{fig:bakboson}.

\begin{figure}[h]
\centerline{\psfig{file=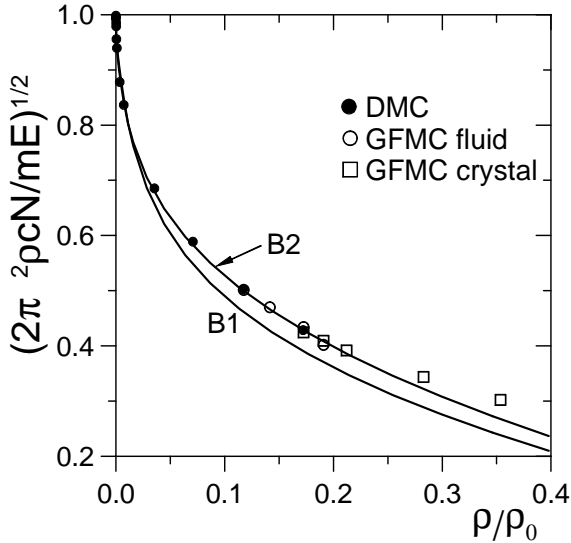,height=3.0in,width=3.0in}}
\caption{The quantity $\protect\epsilon _{0}^{-1/2}=\protect\sqrt{2\protect%
\pi \hbar ^{2}\protect\rho \,c\,N/m\,E}=[1-(\protect\rho /\protect\rho %
_{0})^{1/3}]\protect\sqrt{1+b\,(\protect\rho /\protect\rho _{0})^{1/3}}$ as
a function of $x/x_{0}$ for boson hard sphere systems: B1 and B2 refer to (%
\protect\ref{eps0B}) and (\protect\ref{deps0-1/2}) with $A\simeq 15.7$ and $%
A\simeq 11.9$, respectively. Larger dots are GFMC fluid data and smaller
dots refer to DMC (fluid) calculations.}
\label{fig:bakboson}
\end{figure}

Alternatively, if we allow the residue $A$ to be free one may ask for a
solution minimizing $\sum_{i=1}^{N}[\epsilon
_{0-DMC}^{-1/2}(x_{i}^{DMC})-\epsilon _{0B}^{-1/2}(\alpha
,A,x_{i}^{DMC})]^{2}$ with respect $\alpha $ and $A$, i.e., 
\[
\frac{d}{d\alpha }\sum_{i=1}^{N}[\epsilon
_{0-DMC}^{-1/2}(x_{i}^{DMC})-\epsilon _{0B}^{-1/2}(\alpha
,A,x_{i}^{DMC})]^{2}=-\sum_{i=1}^{N}2[\epsilon
_{0-DMC}^{-1/2}(x_{i}^{DMC})-\epsilon _{0B}^{-1/2}(\alpha ,A,x_{i}^{DMC})]%
\frac{d}{d\alpha }\epsilon _{0B}^{-1/2}(\alpha ,A,x_{i}^{DMC})=0 
\]%
or 
\begin{equation}
\sum_{i=1}^{N}2[\epsilon _{0-DMC}^{-1/2}(x_{i}^{DMC})-\epsilon
_{0B}^{-1/2}(\alpha ,A,x_{i}^{DMC})]\frac{-Y(\alpha ,A,x_{i})x^{2}+(A/4\pi
)^{-1/2}\frac{2}{3}%
x_{rccp}^{1/3}x_{rccp}^{2}[(x/x_{rccp})^{2}-(x/x_{rccp})^{3}]}{%
1-K_{2}\,x^{2}\ln \,x^{2}+\alpha x^{2}}=0  \label{alfamin}
\end{equation}%
as well as of 
\[
\frac{d}{dA}\sum_{i=1}^{N}[\epsilon _{0-DMC}^{-1/2}(x_{i}^{DMC})-\epsilon
_{0B}^{-1/2}(\alpha ,A,x_{i}^{DMC})]^{2}=-\sum_{i=1}^{N}2[\epsilon
_{0-DMC}^{-1/2}(x_{i}^{DMC})-\epsilon _{0B}^{-1/2}(\alpha ,A,x_{i}^{DMC})]%
\frac{d}{dA}\epsilon _{0B}^{-1/2}(\alpha ,A,x_{i}^{DMC})=0 
\]%
or 
\begin{equation}
\sum_{i=1}^{N}2[\epsilon _{0-DMC}^{-1/2}(x_{i}^{DMC})-\epsilon
_{0B}^{-1/2}(\alpha ,A,x_{i}^{DMC})]\frac{(A/4\pi )^{-3/2}\frac{1}{12\pi }%
x_{rccp}^{1/3}(1-K_{2}\,x_{rccp}^{2}\ln \,x_{rccp}^{2}+\alpha
x_{rccp}^{2})[(x/x_{rccp})^{2}-(x/x_{rccp})^{3}]}{1-K_{2}\,x^{2}\ln
\,x^{2}+\alpha x^{2}}=0\,.  \label{Amin}
\end{equation}%
Under the two conditions (\ref{alfamin}) and (\ref{Amin}) we find an optimal 
$A\simeq 11.8715$ and an optimal $\alpha \simeq 169.516$, leading to $\beta
\simeq 124.1$ and $\gamma \simeq -111.296$. This procedure gives the curve
labeled B2 in Fig. \ref{fig:bakboson}. Note that the residue $11.8715$ is
now being associated with the \textit{random} closest close-packing (rccp)
density $0.86\rho _{0}$\ of hard spheres. This value of \ $A$\ is somewhat
smaller than the residue $15.667$\ at \textit{regular} close-packing density 
$\rho _{0}$, though still within the rigorous range stated in (\ref%
{rigorousRange}). 
\begin{figure}[h]
\centerline{\psfig{file=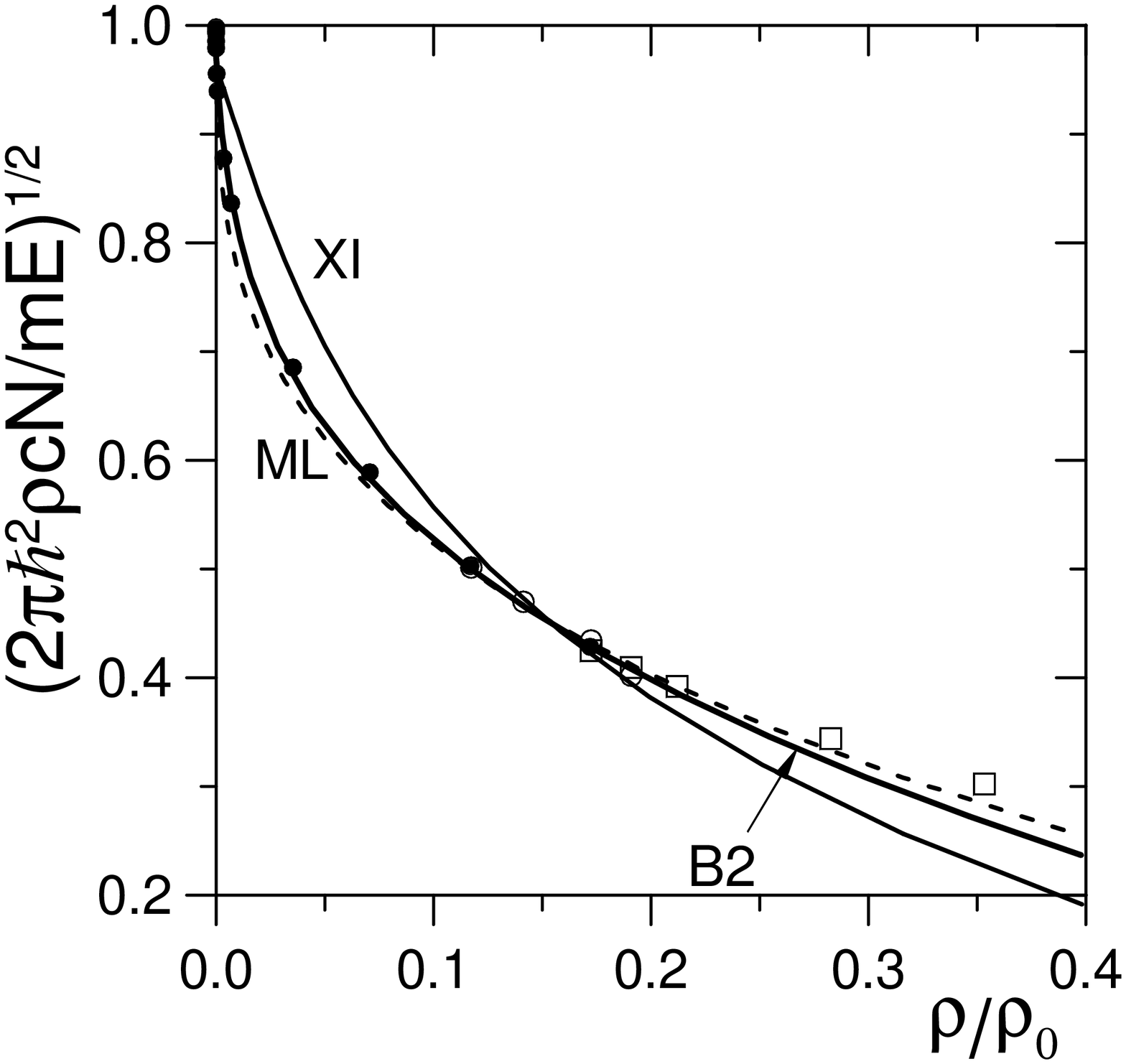,height=3.0in,width=3.0in}}
\caption{The quantity $\protect\epsilon _{0}^{-1/2}=\protect\sqrt{2\protect%
\pi \hbar ^{2}\protect\rho \,c\,N/m\,E}=[1-(\protect\rho /\protect\rho %
_{0})^{1/3}]\protect\sqrt{1+b\,(\protect\rho /\protect\rho _{0})^{1/3}}$ as
a function of $\protect\rho /\protect\rho _{0}$ for the boson hard sphere
system: XI is the fluid branch approximant of Ref. \protect\cite{Annals},
Fig. 2; B2 refers to (\protect\ref{eps0B}) and (\protect\ref{deps0-1/2})
with $A\simeq 11.9$; ML is the modified London formula (\protect\ref{london}%
). Open circles and squares are GFMC data for the fluid and crystalline
branches, respectively, and dots are DMC data points.}
\label{fig:kirw-b1}
\end{figure}
Figure \ref{fig:kirw-b1} compares the previous fluid branch expression XI$%
(x),$ Eq. (17) of Ref.\ \cite{Annals}, with the present extrapolant (\ref%
{eps0B}) labeled B2, both as full curves. The dashed curve is the modified
London (ML) formula (\ref{london}) that connects smoothly with the
crystalline branch. Open circles and squares are GFMC data for fluid and
crystalline branches, respectively. Dots represent DMC \cite{Boronat} data
spanning a wider range of densities in the fluid region than the GFMC data.
The new expression B2 shows dramatically better agreement with DMC data for
intermediate densities, as well as agreeing well with both DMC and GFMC data
around the freezing transition mentioned in Table I of Ref.~\cite{GFMC}.
Fig.~3 is an enlargement of Fig.~2 at low densities to show the remarkable
agreement of B2 with the DMC\ data.

\section{Fermion hard-sphere fluid branch}

The ground-state energy per particle for fermion hard-sphere fluids (\ref%
{fer})\ can be written as 
\begin{equation}
E/N\;=\;\frac{3}{5}\frac{\hbar ^{2}k_{F}^{2}}{2m}\,e_{0}(x)\,,\qquad x\equiv
k_{F}c\,  \label{ENF}
\end{equation}%
with 
\begin{eqnarray}
e_{0}(x)\,\equiv 1 &+&C_{1}\,x+C_{2}\,x^{2}+(C_{3}/3+C_{4}/3+C_{5})\,x^{3} 
\nonumber \\
&+&C_{6}\,x^{4}\,\ln x+(C_{7}/3-C_{8}/3+C_{9})\,x^{4}+o(x^{4})\,
\label{fer2}
\end{eqnarray}

for $x\equiv k_{F}c\ll 1$, $\rho \equiv N/\Omega =\nu k_{F}^{3}/6\pi ^{2}$
being the number of fermions $N$ in the enclosed volume $\Omega .$ We shall
examine both $\nu =2$ (corresponding to liquid $^{3}$He and neutron matter)
and $\nu =4$ (corresponding to nuclear matter). 

\begin{figure}[h]
\centerline{\psfig{file=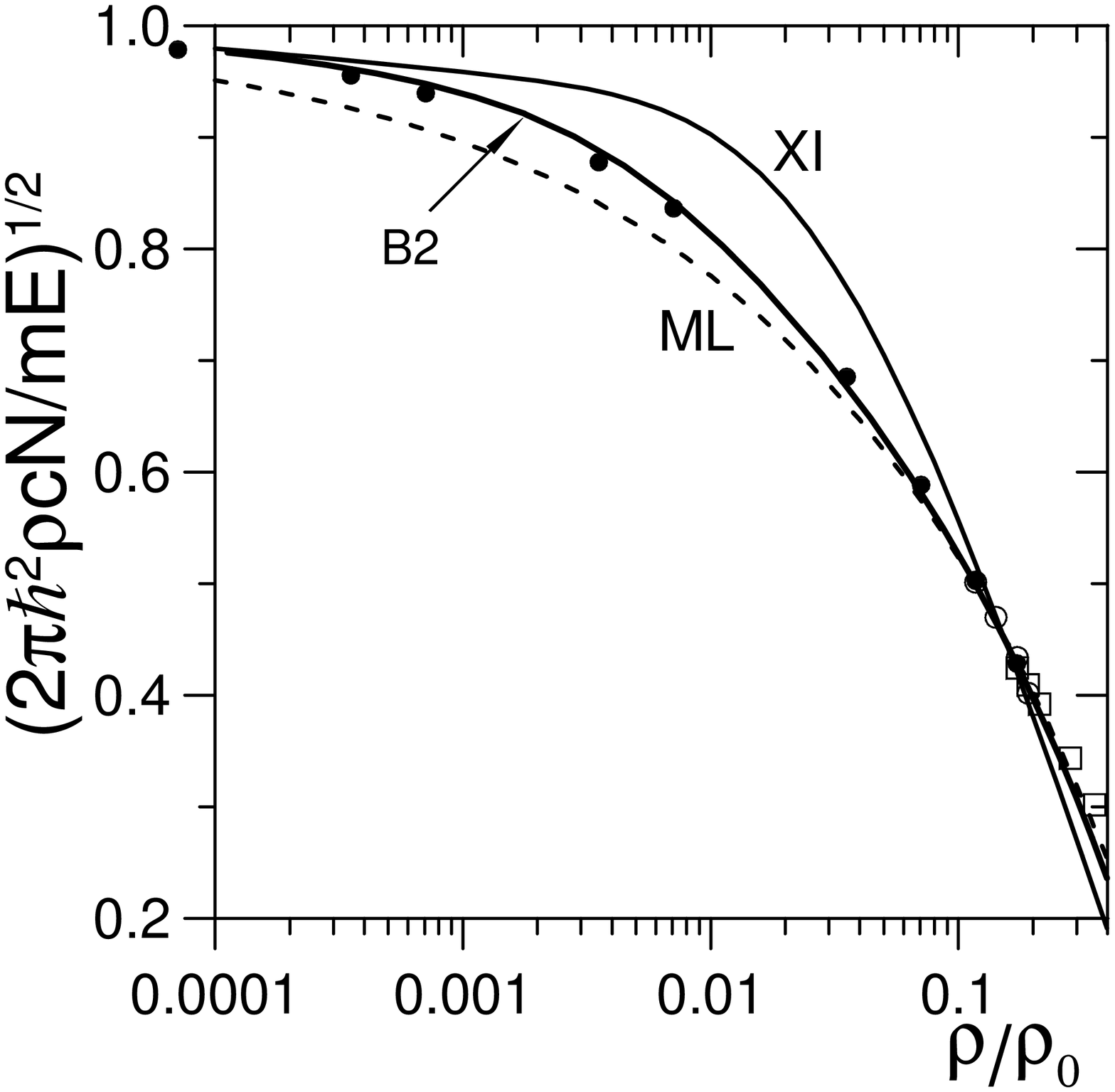,height=3.0in,width=3.0in}} \vspace{%
-0.30cm}
\caption{Enlargement of Fig. \protect\ref{fig:kirw-b1} at low densities.}
\label{fig:kirw-b12}
\end{figure}

\subsection{Fermions with $\protect\nu =2$}

\begin{figure}[h]
\centerline{\psfig{file=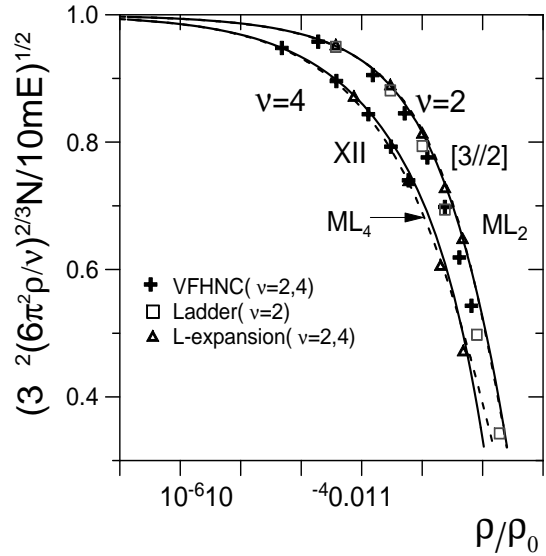,height=3.0in,width=3.0in}} 
\caption{The expression (\protect\ref{formforFig2}) as function of $\protect%
\rho /\protect\rho _{0}$ for fermion hard spheres with $\protect\nu =2$
labeled [3//2] and with $\protect\nu =4$ labeled XII (full curves). Dashed
curves are the corresponding modified London ML$_{\protect\nu }$ formulae,
but note that the ML$_{2}$ dashed curve almost coincides with the full curve
[3//2].}
\label{fig:kirw-f1}
\end{figure}

For $\nu =2$, $C_{6}=0$ \cite{fermion} so that (\ref{fer2}) simplifies to
the pure power series 
\begin{equation}
e_{0}(x)\,=\,1+D_{1}\,x+D_{2}\,x^{2}+D_{3}\,x^{3}+D_{4}\,x^{4}+o(x^{4})
\label{fer3}
\end{equation}%
where the $C_{i}$'s have been determined in terms of the $D_{i}$'s. As in
the boson case, instead of $e_{0}(x)$ we consider the series 
\begin{equation}
e_{0}^{-1/2}(x)=1+F_{1}\,x+F_{2}\,x^{2}+F_{3}\,x^{3}+F_{4}\,x^{4}+F_{5}%
\,x^{5}+o(x^{5})\,  \label{eps0nu2}
\end{equation}%
where the $F_{i}$'s depend algebraically on the $D_{i}$'s in a simple
manner, $F_{5}$ being unknown. Values of $D_{i}$ and $F_{i}$ are given in
Table~\ref{tab:Nu2Coef}. We use this simple power series to construct the
usual Pad\'{e} extrapolants. The approximants to (\ref{eps0nu2}) with four
terms beyond the trivial unity were analyzed in Ref.~\cite{Hu} where it was
concluded that the best approximant was the Pad\'{e} $[0/4](x)$. However,
this function does not have a zero in the region of physical interest, i.e., 
$0\leq \rho /\rho _{0}\leq 1$, which implies $0\leq x\equiv k_{F}c\leq 3.47$
since $\rho =k_{F}^{3}/3\pi ^{2}$. Accordingly, the energy does not manifest
a close-packing density as it should. This deficiency made it advisable to
introduce the 
fifth term $F_{5}\,x^{5}$\ in (\ref{eps0nu2}). Although in Fig. 1 of Ref. 
\cite{puertorico} only five of the six \textit{two-point }Pad\'{e}
approximants $[L//M](x)$ with $L+M=5$, are shown, all six approximants were
analyzed here to adjust $F_{5}$ so as to ensure a zero associated with a
random close-packing in the physical region. The approximant $\epsilon
_{0}(x)$ and the position of its zero were chosen in such way that the QTPT
applied in Ref.~\cite{Hu} to calculate the ground-state energy of $^{3}$He
with the Aziz interatomic potential \cite{Aziz} reproduces the corresponding
GFMC \cite{Panoff} data. (In this treatment, the Aziz potential was
decomposed via the well-known Barker-Henderson (BH) \cite{BH} scheme as
described in Ref.~\cite{Hu}.) Eventually, the best extrapolant was found to
be the \textit{two-point }Pad\'{e} approximant 
\begin{equation}
e_{0}^{-1/2}(x)\doteq \;[3//2](x)\,\equiv \frac{%
N_{0}+N_{1}x+N_{2}x^{2}+N_{3}x^{3}}{M_{0}+M_{1}x+M_{2}x^{2}}\equiv \epsilon
_{0}^{-1/2}(x)  \label{3//2}
\end{equation}%
where 
\begin{eqnarray}
N_{0} &=&F_{2}F_{4}-F_{3}^{2}  \nonumber \\
N_{1} &=&F_{4}(F_{3}+F_{1}F_{2})-F_{2}F_{5}-F_{1}F_{3}^{2}  \nonumber \\
N_{2}
&=&(F_{3}-F_{1}F_{2})F_{5}-F_{4}^{2}+(F_{1}F_{3}+F_{2}^{2})F_{4}-F_{2}F_{3}^{2}
\nonumber \\
N_{3}
&=&(F_{1}F_{3}-F_{2}^{2})F_{5}-F_{1}F_{4}^{2}+2F_{2}F_{3}F_{4}-F_{3}^{3} 
\nonumber \\
M_{0} &=&F_{2}F_{4}-F_{3}^{2}  \nonumber \\
M_{1} &=&F_{3}F_{4}-F_{2}F_{5}  \nonumber \\
M_{2} &=&F_{3}F_{5}-F_{4}^{2}.  \nonumber
\end{eqnarray}%
\begin{table}[b]
\begin{center}
\begin{tabular}{||c|c|c|c|c|c||}
\hline\hline
$\nu=2$ & $i=$1 & 2 & 3 & 4 & 5 \\ \hline
$D_i$ & 0.353678 & 0.185537 & 0.384145 & -0.024700 & ``-0.265544" \\ \hline
$F_i$ & -0.176833 & -0.045863 & -0.156677 & 0.109672 & ``0.130830" \\ 
\hline\hline
\end{tabular}
\\[0pt]
\end{center}
\caption{Coefficients $D_{i}$ and $F_{i}$ for $\protect\nu =2$ \ appearing
in (\protect\ref{fer3}) and (\protect\ref{eps0nu2}), respectively. Numbers
in quotation marks were determined as indicated in text.}
\label{tab:Nu2Coef}
\end{table}
The extrapolant (\ref{3//2})\ satisfies $[3//2](x=3.13)=0$. Hence the
ground-state energy per fermion for $\nu =2$ becomes 
\begin{equation}
E/N\;\doteq \;\frac{3}{5}\frac{\hbar ^{2}k_{F}^{2}}{2m}\,\{[3//2](x)\}^{-2}\,
\label{efer2}
\end{equation}%
with a random closest close-packing density $\rho _{rccp}/\rho _{0}=0.732$
only 15 \% smaller than the empirical \cite{ScottKilgour}\ value $\rho
_{rccp}/\rho _{0}\simeq 0.86$. The coefficient $F_{5}$ is listed in Table~%
\ref{tab:Nu2Coef} in quotation marks. In Fig.~\ref{fig:kirw-f1} we show the
expression 
\begin{eqnarray}
&&\epsilon _{0}^{-1/2}=[3\hbar ^{2}(6\pi ^{2}\rho /\nu
)^{2/3}N/10mE]^{1/2}=1+  \nonumber \\
&&[{20\pi (\nu -1)}/{3\nu }]({2^{1/4}\nu }/{6\pi ^{2}})^{2/3}\{[(\rho /\rho
_{0})^{-1/3}-1]^{2}[(\rho /\rho _{0})^{-1/3}-b(\nu )](\rho /\rho
_{0})^{2/3}\}^{-1}  \label{formforFig2}
\end{eqnarray}%
as a function of $\rho /\rho _{0}$ for fermion hard spheres. Here $b(\nu )$
is as defined in (\ref{bnu}). For $\nu =2$ the fluid branch [3//2] (full
curve) given by (\ref{3//2}) is close to the Ladder 
\cite{Ex.Ladd} (open squares), the variational Fermi hypernetted chain
(VFHNC) \cite{VFHNC} (plus-sign marks), and the so-called L-expansion data 
\cite{L-expansion,L-expansion1} (open triangles). Fig.~\ref{fig:kirw-f1}
shows good agreement over the entire range of available data. 

In order to improve the many-fermion ground-state energy equation of state
we include the next term in (\ref{eps0nu2}), i.e., $F_{6}\,x^{6}$, which is
then used to generate all Pad\'{e} approximants of order six to the series \ 
$\epsilon _{0}^{-1/2}(x)$. The lack of a logarithmic term $x^{4}\ln x$ is
due to the Pauli principle \cite{fermion}. Such a term arises when there are 
\textit{three} independent hole lines. But for $\nu =2$ there can be at most 
\textit{two} lines of the same spin. Thus the Pauli principle reduces the
size of the term by a factor of the density. We thus expect the first such
term for $\nu =2$ to be $O(x^{7}\ln x)$. The unknown coefficients $F_{6}$
and\ $F_{5}$ are determined from the two double-pole conditions (\ref{C1})
and (\ref{C2}), which become 
\begin{equation}
\epsilon _{0}^{-1/2}(x)\qquad \smash {\
\mathop{\relbar\joinrel\longrightarrow}\limits_{\rho \to \rho_{rccp}}\ \ }%
\qquad (1-x/x_{rccp})[5A/3(3\pi ^{2})^{2/3}]^{-1/2}  \label{C1F2}
\end{equation}%
and 
\begin{equation}
\epsilon _{0}(x)+\frac{x}{2}\frac{d\epsilon _{0}(x)}{dx}\qquad \smash {\
\mathop{\relbar\joinrel\longrightarrow}\limits_{\rho \to \rho_{rccp}}\ \ }%
\qquad \frac{5A/3(3\pi ^{2})^{2/3}}{(1-x/x_{rccp})^{3}}  \label{C2F2}
\end{equation}%
with $A\simeq 15.667$.

\begin{figure}[h]
\centerline{\psfig{file=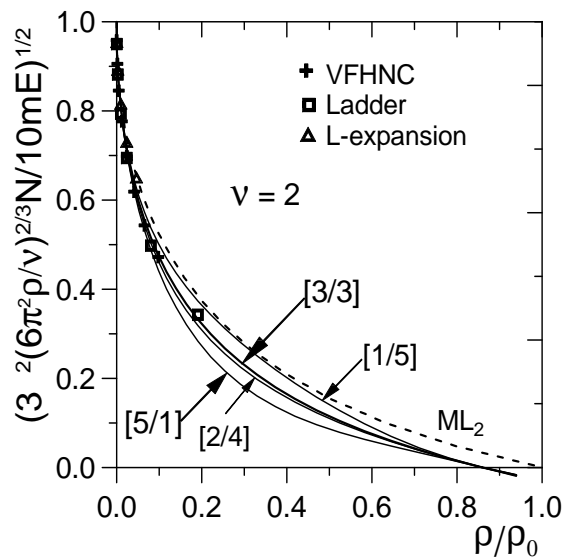,height=3.0in,width=3.0in}}
\caption{Improved extrapolants for the many-fermion hard-sphere gas with $%
\protect\nu =2$.}
\label{fig:Ba-MA-Fer4}
\end{figure}
\begin{figure}[h]
\centerline{\psfig{file=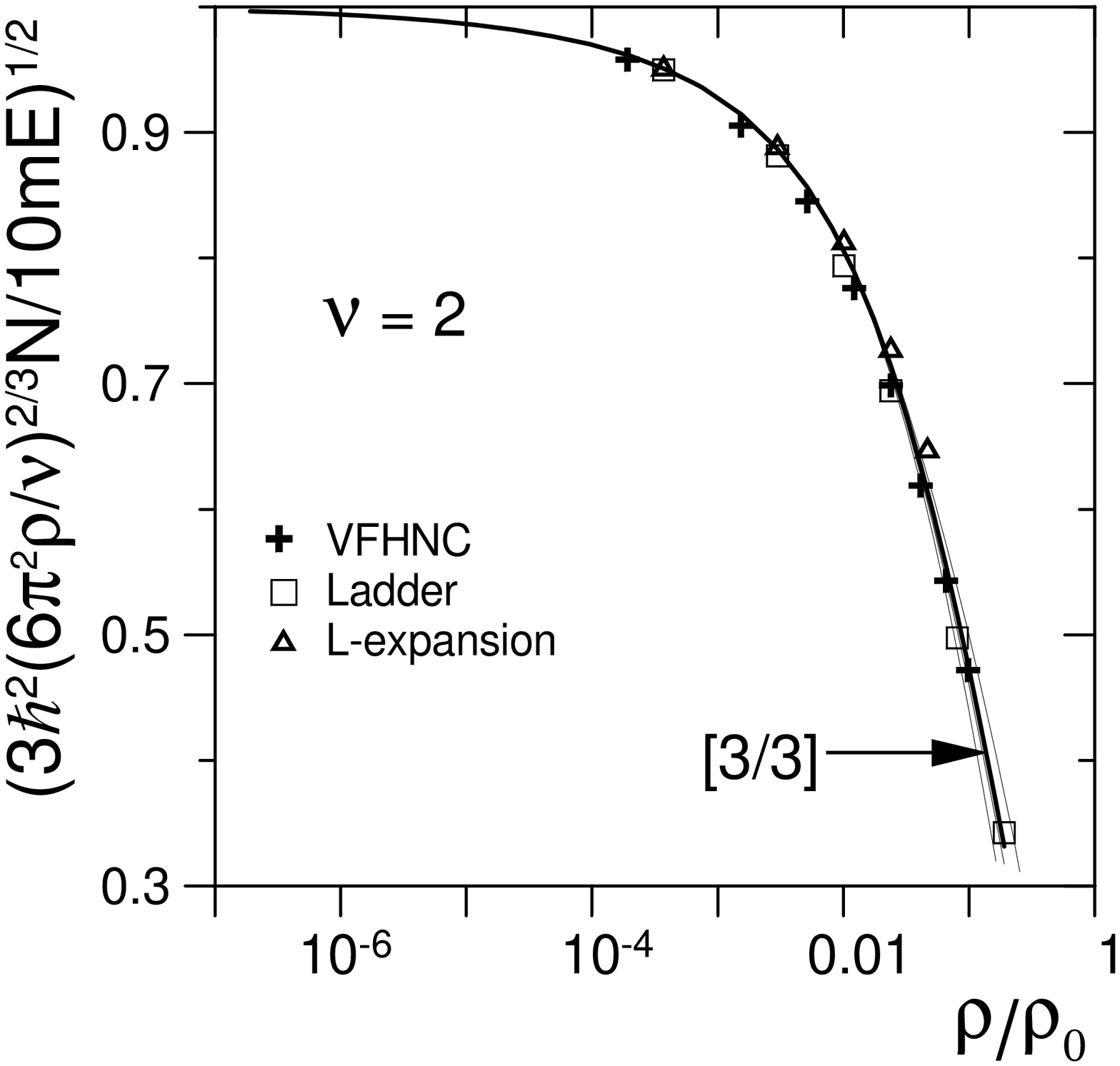,height=3.0in,width=3.0in}}
\caption{Enlargement of Fig. \protect\ref{fig:Ba-MA-Fer4} at low densities.}
\label{fig:Ba-MA-Fer3}
\end{figure}
\begin{figure}[h]
\centerline{\psfig{file=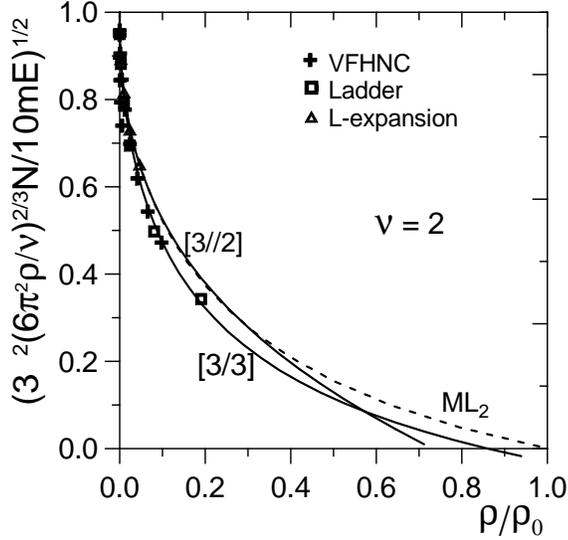,height=3.0in,width=3.0in}} 
\caption{Comparison of quantity (\protect\ref{formforFig2}) as function of $%
\protect\rho /\protect\rho _{0}$ for many-fermion hard spheres with $\protect%
\nu =2$, for the the previously best approximant $[3//2](x)$ \protect\cite%
{Annals} and the new improved one $[3/3](x)$, full curves. Dashed curve is
modified London formula.}
\label{fig:Ba-MA-nu2}
\end{figure}

For each Pad\'{e} approximant of order six we determined $F_{6}$ and\ $F_{5}$
as shown in Table~\ref{tab:F5yF6Nu2Coef}. Approximants [4/2] and [0/6] did
not exhibit the double-pole conditions. The other four approximants are
plotted in Fig.~\ref{fig:Ba-MA-Fer4} together with the Ladder \cite{Ex.Ladd}
(open squares), the variational Fermi hypernetted chain (VFHNC) \cite{VFHNC}
(plus-sign marks), and the L-expansion \cite{L-expansion} (open triangles)
data for $\nu =2$, from which we conclude that the approximant $[3/3](x)$ is
the best. Fig. \ref{fig:Ba-MA-Fer3} is a semi-log enlargement of Fig. \ref%
{fig:Ba-MA-Fer4}. In Fig. \ref{fig:Ba-MA-nu2} we compare both the new
improved expression $[3/3](x)$ and the previous best energy expression, i.e,
the two-point Pad\'{e} approximant $[3//2](x)$ reported in Ref. \cite{Annals}
and supported by Ladder, VFHNC and L-expansion data. 
\begin{table}[htb]
\begin{center}
\begin{tabular}{||c|c|c||}
\hline\hline
Pad\'e & F5 & F6 \\ \hline
[5/1] & -0.0272548 & 0.0038205 \\ \hline
[4/2] & -.20 & no solution \\ \hline
[3/3] & -0.0130625 & 0.0039120 \\ \hline
[2/4] & -0.0395076 & 0.0415222 \\ \hline
[1/5] & -0.0115902 & 0.01887153 \\ \hline
[0/6] & -0.1276 & no solution \\ \hline\hline
\end{tabular}
\\[0pt]
\end{center}
\caption{The $F_{5}$ and $F_{6}$ coefficients for $\protect\nu =2$ that
follow from conditions (\protect\ref{C1F2}) and (\protect\ref{C2F2}) for all
sixth-order Pad\'{e} approximants with residue $A\simeq 15.667$ and random
closest close-packing density $\protect\rho _{rccp}\equiv 0.86\protect\rho %
_{0}$.}
\label{tab:F5yF6Nu2Coef}
\end{table}

\pagebreak \pagebreak

\subsection{Fermions with{\ $\protect\nu =4$ }}

For fermions with $\nu =4$ (\ref{fer2}) becomes 
\begin{equation}
e_{0}(x)\,=\,1+D_{1}\,x+D_{2}\,x^{2}+D_{3}\,x^{3}+D_{4}\,x^{4}\,\ln
x+D_{5}\,x^{4}+o(x^{4})  \label{fer4}
\end{equation}%
for $x\equiv k_{F}c\ll 1$ and we recall that $\rho =\nu k_{F}^{3}/6\pi ^{2}$%
. As for bosons or for fermions with $\nu =2$, we analyze 
\begin{equation}
e_{0}^{-1/2}(x)=1+F_{1}\,x+F_{2}\,x^{2}+F_{3}\,x^{3}+F_{4}\,x^{4}\,\ln
x+F_{5}\,x^{4}+o(x^{4})  \label{eps0nu4}
\end{equation}%
with all $F_{i}$ $(i=1,2,3,4)$\ known. Values of $D_{i}$ and $F_{i}$ are
given in Table~\ref{tab:Nu4Coef}. Unlike the $\nu =2$ case, this series is
not a pure power series as it contains logarithmic terms. Its so-called
\textquotedblleft tailing\textquotedblright\ \cite{tailing}\ approximants
are given in Table III of Ref.~\cite{formas}. Of all the possible
approximants using only the known coefficients, only the forms II and XII
are free from 
flaws and have residues within the bounds (\ref{rigorousRange}). Of these
two forms, II has a residue less than that predicted in Ref.~\cite{Hubbard}.
Hence we chose form XII, which is plotted in Fig.~\ref{fig:kirw-f1} as the
full curve labeled XII. 
\begin{table}[tbp]
\begin{center}
\begin{tabular}{||c|c|c|c|c||}
\hline\hline
$\nu=4$ & $i=$1 & 2 & 3 & 4 \\ \hline
$D_i$ & 1.061033 & 0.556610 & 1.300620 & -1.408598 \\ \hline
$F_i$ & -0.530517 & 0.143867 & -0.5806558 & -0.704299 \\ \hline\hline
\end{tabular}
\\[0pt]
\end{center}
\caption{Coefficients $D_{i}$ and $F_{i}$ for $\protect\nu =4$ appearing in~(%
\protect\ref{fer4}) and (\protect\ref{eps0nu4}), respectively.}
\label{tab:Nu4Coef}
\end{table}

In this case $E/N$ can be written as 
\begin{equation}
E/N=\frac{3\hbar ^{2}k_{F}^{2}}{10m}\epsilon _{0}(x)\,
\end{equation}%
where the series (\ref{eps0nu4}) is represented as 
\[
e_{0}^{-1/2}(x)\doteq \mbox{XII}(x)\equiv \frac{%
1+(F_{1}-F_{3}/F_{2})x+(F_{2}-F_{1}F_{3}/F_{2})x^{2}}{%
1-(F_{3}/F_{2})x-F_{4}x^{4}\ln x}\equiv \epsilon _{0}^{-1/2}(x). 
\]
We also plot the corresponding VFHNC data (plus-sign marks) and L-expansion
data (open triangles). In terms of energy, our results are slightly below
the VFHNC points, with agreement improving at lower densities. On the other
hand, the XII approximant lies just above the L-expansion data over the
range of densities where data are available.

In order to improve the $\nu =4$ many-fermion hard-sphere ground-state
energy equation of state, the energy series (\ref{fer}) was written as 
\begin{eqnarray}
{\frac{E}{N}}-{\frac{3}{5}}{\frac{{\hbar ^{2}k_{F}^{2}}}{{2m}}}={\frac{3}{5}}%
{\frac{{\hbar ^{2}k_{F}^{3}c}}{{2m}}}e_{0}(x)= &&{\frac{3}{5}}{\frac{{\hbar
^{2}k_{F}^{3}c}}{{2m}}}[D_{1}+D_{2}x+D_{3}x^{2}+D_{4}x^{3}\ln x+D_{5}x^{3} 
\nonumber \\
&&+D_{6}x^{4}\ln x+D_{7}x^{4}+\cdots ]
\end{eqnarray}
where $x=k_{F}c$ and $\rho =\nu k_{F}^{3}/6\pi ^{2}$. The suggested
representation for $e_{0}(x)$ here is 
\begin{equation}
e_{0}(x)=D_{1}+D_{2}x+D_{3}x^{2}+D_{4}x^{3}\ln x+D_{5}x^{3}+D_{6}x^{4}\ln
x+D_{7}x^{4}+\cdots  \label{DnewCoef}
\end{equation}
which leads to 
\begin{equation}
e_{0}(x)^{-1/2}=F_{1}+F_{2}x+F_{3}x^{2}+F_{4}x^{3}\ln
x+F_{5}x^{3}+F_{6}x^{4}\ln x+F_{7}x^{4}+\cdots  \label{FnewCoef}
\end{equation}
with $D_{1}$ to $D_{4}$ known and equal to the values given in the Table~\ref%
{tab:Nu4Coef}. The coefficients $F_{1}$ to $F_{4}$ are different from those
in Table~\ref{tab:Nu4Coef}, but they are derived simply from the $D_{i}$'s
and so are also known. They are 
\[
F_{1}=1/\sqrt{D_{1}};\quad F_{2}=-D_{2}/2D_{1}^{3/2};\quad F_{3}=(3{D_{2}}%
^{2}-4D_{1}D_{3})/8D_{1}^{5/2};\quad F_{4}=-8D_{1}^{2}D_{4}/16D_{1}^{7/2}; 
\]
\[
F_{5}=(-5D_{2}^{3}+12D_{1}D_{2}D_{3}-8D_{1}^{2}D_{5})/16D_{1}^{7/2};\quad
F_{6}=32D_{1}^{2}(3D_{2}D_{4}-2D_{1}D_{6})/128D_{1}^{9/2}. 
\]

We have also investigated the representation 
\[
\epsilon _{0}^{-1/2}(x)={\frac{{F_{1}+F_{2}x+F_{3}x^{2}+bx^{3}}}{{%
1-(F_{4}/F_{1})x^{3}\ln x+ax^{3}}}} 
\]
for which the two double-pole conditions (\ref{C1}) and (\ref{C2}) imply
that 
\[
b=-x_{rccp}^{-3}[F_{1}+F_{2}x_{rccp}+F_{3}x_{rccp}^{2}] 
\]
and 
\[
F_{2}+2F_{3}x_{rccp}+3bx_{rccp}^{2}=-\left( {\frac{3}{{5x_{rccp}}}}\right) ^{%
\frac{1}{2}}\left( {\frac{3}{{\pi \nu }}}\right) ^{\frac{1}{3}}\left[ 1-{%
\frac{{F_{4}}}{{F_{1}}}}x_{rccp}\ln x_{rccp}+ax_{rccp}^{3}\right] . 
\]
The values of $a$ and $b$ so determined are $-0.0924883$ and 0.171942,
respectively. This representation is unsatisfactory because it has what
applied mathematicians call a \textquotedblleft defect.\textquotedblright\
Unfortunately it is in the physical region $0<x<x_{rccp}$. The problem is
not uncommon and stems from a pole and a zero lying very close to each other.

\section{Conclusions}

Based on known terms of field-theoretic perturbative low-density expansions
we have constructed closed-form analytical expressions as functions of
particle density using Pad\'{e} and other approximants for the energy per
particle of the fluid branches of both many-boson and many-fermion quantum
hard-sphere systems. Improvements with respect to previous work (notably but
not exclusively that of Ref.~\cite{Annals}) have been achieved by assuming
i) that the classical random closest close-packing hard-sphere densities are
the ultimate fluid densities at which the energy diverges with a
second-order pole and ii) proposing and imposing a value for the residue at
the pole that is the same for either bosons or fermions as closest
close-packing is approached and the hard spheres become distinguisable.
Implementing these two conditions and taking advantage of recent diffusion
Monte Carlo simulation data has allowed us to incorporate an additional term
in the low-density expansion beyond that employed in Ref.~\cite{Annals}. The
resulting determination of the best approximants has produced decidedly
improved results for bosons as well as for two-component fermions, but not
for four-component fermions. \newline

\noindent\textbf{ACKNOWLEDGMENTS}

MAS thanks Washington University for hospitality during a sabbatical year
and support as an Edwin T.~Jaynes Visiting Professor. MdeLl thanks the NSF
(USA) for partial support through grant INT-0336343 made to the \textit{%
Consortium of the Americas for Interdisciplinary Science}, University of New
Mexico, Albuquerque, NM, USA.\ We also acknowledge research support from
UNAM-DGAPA-PAPIIT (Mexico) grants IN106401 and IN108205, from CONACyT
(Mexico) grants 27828-E and 41302-F, and from NSF (USA) grant~PHY-0140316.
This work was also supported in part by the U.S. Department of Energy at the
Los Alamos National Laboratory under Contract DE-AC 52-06NA25396.

\pagebreak

\end{document}